\documentclass[10pt, conference]{IEEEtran}

\newcommand{\coolname}{\textrm{Contego}\xspace}

\title{\coolname: An Adaptive Framework for Integrating Security Tasks in Real-Time Systems}

\usepackage[utf8]{inputenc}
\usepackage{amsmath}
\usepackage{amsfonts}

\usepackage{amsthm}

\usepackage{float}

\usepackage{graphicx}
\graphicspath{{Figures/}}
\usepackage{epstopdf}

\usepackage{epsfig}
\usepackage{url}
\usepackage{comment}
\usepackage{cite}
\usepackage{graphics}
\usepackage{soul} 
\usepackage{multirow}

\usepackage{todonotes} 

\usepackage{graphicx, array, blindtext} 

\usepackage{graphicx}

\usepackage{wrapfig}

\usepackage{algorithm}  
\usepackage{algorithmic}

\usepackage{listings}
\usepackage{psfrag,wrapfig}

\usepackage{enumerate}
\usepackage{paralist} 
\usepackage{subfiles} 

\newcommand{\eg}{{\it e.g.,}\xspace}
\newcommand{\viz}{{\it viz.,}\xspace}

\newcommand{\ie}{{\it i.e.,}\xspace}
\newcommand{\etc}{{\it etc.}}
\newcommand{\ci}{{\it (i) }}
\newcommand{\cii}{{\it (ii) }}

\newcommand{\ca}{{\it (a) }}
\newcommand{\cb}{{\it (b) }}

\soulfont{\eg}{0}
\soulfont{\ie}{0}
\soulfont{\etc}{0}
\soulfont{\viz}{0}
\soulfont{\coolname}{0}
\soulfont{\ave}{0}
\soulfont{\pve}{0}

\usepackage{multirow}

\newcounter{myoptimizationproblemctr}
\newenvironment{myoptimizationproblem}{
   \bigskip\noindent
   \refstepcounter{myoptimizationproblemctr}
   $(\mathbf{P\themyoptimizationproblemctr})$
}

 \newtheorem{remark}{Remark}

\usepackage{array}
\usepackage{ragged2e}
\newcolumntype{P}[1]{>{\RaggedRight\hspace{0pt}}p{#1}}

\usepackage{relsize} 

\newcommand{\pve}{\xspace{\relsize{-1.20}\textsc{PASSIVE}}\xspace}

\newcommand{\ave}{\xspace{\relsize{-1.20}\textsc{ACTIVE}}\xspace}

\begin{document}






  \author{\IEEEauthorblockN{Monowar Hasan\IEEEauthorrefmark{1}, Sibin Mohan\IEEEauthorrefmark{1},  Rodolfo Pellizzoni\IEEEauthorrefmark{2} and Rakesh B. Bobba\IEEEauthorrefmark{3}} \IEEEauthorblockA{\IEEEauthorrefmark{1}Dept. of Computer Science, University of Illinois at Urbana-Champaign, Urbana, IL, USA}
  \IEEEauthorblockA{\IEEEauthorrefmark{2}Dept. of Electrical and Computer Engineering, University of Waterloo, Ontario, Canada}
  \IEEEauthorblockA{\IEEEauthorrefmark{3}School of Electrical Engineering and Computer Science, Oregon State University, Corvallis, OR, USA}
  Email: \{\IEEEauthorrefmark{1}mhasan11, \IEEEauthorrefmark{1}sibin\}@illinois.edu,
  \IEEEauthorrefmark{2}rodolfo.pellizzoni@uwaterloo.ca,
 \IEEEauthorrefmark{3}rakesh.bobba@oregonstate.edu}

\maketitle

 \thispagestyle{plain}
 \pagestyle{plain}


\begin{abstract}


Embedded real-time systems (RTS) are pervasive. Many modern RTS are exposed to unknown security flaws, and threats to RTS are growing in both number and sophistication. However, until recently, cyber-security considerations were an afterthought in the design of such systems.  
Any security mechanisms integrated into RTS must \ca \textit{co-exist} with the real-time tasks in the system and \cb operate \textit{without} impacting the timing and safety constraints of the control logic. We introduce \coolname, an approach to integrating security tasks into RTS without affecting temporal requirements. \coolname is specifically designed for \textit{legacy} systems, \viz the real-time control systems in which major alterations of the system parameters for constituent tasks is not always feasible.
\coolname combines the concept of \textit{opportunistic execution} with hierarchical scheduling to maintain compatibility with legacy systems while still providing flexibility by allowing security tasks to operate in different \textit{modes}. We also define a metric to measure the effectiveness of such integration. 
We evaluate \coolname using synthetic workloads as well as with an implementation on a realistic embedded platform (an open-source ARM CPU running real-time Linux).

\end{abstract}



\section{Introduction}

Embedded real-time systems (RTS) are used to monitor and control physical
systems and processes in many domains, \eg manned and unmanned vehicles
including aircraft, spacecraft, unmanned aerial vehicles (UAVs), submarines and
self-driving cars, critical infrastructures like the electric grid and process control systems in industrial plants, to name just a few.	
They rely on a variety of inputs for 
correct operation and have to meet stringent safety and timing requirements.
Failures in RTS can have catastrophic consequences for the
environment, the system, and/or human safety \cite{abrams2008malicious, checkoway2011comprehensive}. 

Traditionally, RTS were designed using proprietary protocols, platforms and software and were
not connected to the rest of the world, \ie they were air gapped. As a result cyber-security
was not a design priority in such systems.
However, the drive towards remote monitoring and control
facilitated by the growth of the Internet, the rise in the use of
commercial-off-the-shelf (COTS) components, standardized communication protocols
and the high value of these systems to adversaries have been challenging the status
quo. While safety and fault-tolerance have long been important design
considerations in such systems, traditional fault-tolerance techniques that
were designed to counter and survive random or accidental faults are not
sufficient to deal with cyber-attacks orchestrated by an intelligent and
capable adversary. A number of high-profile attacks on real systems, \eg Stuxnet \cite{stuxnet} and attack demonstrations by researchers on
automobiles \cite{ris_rts_1, checkoway2011comprehensive} and medical devices
\cite{security_medical} have shown that the threat is real. 

Given the
increasing cyber-security risks, it is essential to have a layered
defense and integrate resilience against such attacks into the design of
controllers and actuators (\ie embedded RTS). It is
also critical to retrofit existing controllers and actuators with protection,
detection, survival and recovery mechanisms. However, \textit{any security mechanisms have to co-exist with real-time tasks in
the system and have to operate without impacting the timing and safety
constraints of the control logic}. This creates an apparent tension between
security requirements (\eg having enough cycles for effective monitoring and 
detection) and the timing and safety requirements. For example, how often and
how long should a monitoring and detection task run to be effective but not
interfere with real-time control or other safety-critical tasks? While this
tension could potentially be addressed for newer systems at design time, it
is especially challenging in the retrofitting of \textit{legacy} systems for which the control
tasks are already in place and perhaps \textit{cannot be modified}. 
Another challenge
is to ensure that an adversary
cannot easily evade such mechanisms. Further, the deterministic nature of task
schedules in RTS may provide attackers with known windows of opportunity
in which they can run undetected \cite{cy_side_channel,taskshuffler}. 

In order to integrate security mechanisms into RTS, performance criteria such as \textit{frequency of monitoring} and \textit{responsiveness} must be considered. 
For example, security tasks\footnote{We use the terms \textit{security tasks}, \textit{intrusion detection tasks} and \textit{monitoring tasks} interchangeably throughout the paper.} may need to be executed quite frequently to provide good protection. If the interval between consecutive monitoring events is too large, then an attacker may remain undetected and cause harm to the system between two invocations of the security task. In contrast, if the security tasks are executed very frequently, it may impact the schedulability (\ie timely execution) of the real-time tasks \cite{sibin_deeply}. In some circumstances a security task may need to complete with less interference (\eg better responsiveness) from higher-priority real-time tasks. As an example, consider the scenario in which a security breach is suspected and a security task may be required to perform more fine-grained checking instead of waiting for its next  execution slot. At the same time, the scheduling policy needs to ensure that the system does not violate real-time constraints for critical, high-priority control tasks.

Our focus in this work is on \textit{retrofitting security mechanisms into legacy RTS}, for which modification of existing real-time tasks' parameters (such as run-times, period, task execution order, \etc) is not always feasible. In contrast to existing mechanisms \cite{slack_cornell, securecore}, the proposed method does \textit{not} require any architectural modifications and hence is particularity suitable for systems designed using COTS components. The framework developed in this paper is based on our earlier work \cite{mhasan_rtss16} 
in which we proposed to incorporate monitoring and detection mechanisms by implementing them as separate \textit{sporadic tasks} and executing them \textit{opportunistically}, that is, with the lowest priority so that real-time tasks are not affected.
However, if the security tasks always execute with lowest priority, they suffer more interference (\ie preemption from high-priority real-time tasks) and the consequent longer detection time (due to poor response time) will make the security mechanisms less effective. In order to provide \textit{better responsiveness} and increase the effectiveness of monitoring and detection mechanisms,  
we now propose a multi-mode framework called \coolname\footnote{A preliminary version \cite{mhasan_certs16} of this work was presented at a workshop without published proceedings.}. For the most part, \coolname executes in a \pve mode with opportunistic execution of intrusion detection tasks as before~\cite{mhasan_rtss16}. However, \coolname will \textit{switch to an \ave mode of operation} to perform additional checks as needed (\eg fine-grained analysis, used as an example in Section \ref{subsec:implementation_BBB}). This \ave mode potentially executes with higher priority, while ensuring the schedulability of real-time tasks. Thus \coolname subsumes the approach in our earlier work~\cite{mhasan_rtss16} and provides faster detection.  



The contributions of this paper can be summarized as follows:

\begin{itemize}
\item We introduce \coolname, an extensible framework to integrate security tasks into legacy RTS (Section \ref{sec:system_model}). 

\item \coolname allows the security tasks to execute with minimal perturbation of the scheduling order of the real-time tasks while guaranteeing their timing constraints (Sections \ref{sec:sec_server}--\ref{sec:algo}). The proposed method can adapt to changes due to malicious activities by switching its mode of operation. 

\item We propose a metric to measure the security posture of the system in terms of frequency of execution (Section \ref{sec:period_adapt}).

\item We evaluate the schedulability and security of the proposed approach using a range of synthetic task sets and a prototype implementation on an ARM-based development board with real-time Linux (Section \ref{sec:evaluation}).



\end{itemize}


\section{Security and System Model} \label{sec:system_model}



\subsection{Attack Model}
\label{subsec:sec_model}

RTS face threats in various forms, depending on the system and the goals of an adversary. For example, adversaries may insert, eavesdrop on or modify messages exchanged by system components, may manipulate the processing of sensor inputs and actuator commands  and/or could try to modify the control flow of the system \cite{securecore}. Further, rather than try to crash the system aggressively, an intruder in reconnaissance mode may want to monitor the system behavior and gather information for later use. 
For instance, an intruder may utilize side-channels to monitor the system behavior and infer system information (\eg  hardware/software architecture, user tasks and thermal profiles, \etc) that may eventually help maximize the impact  of an attack \cite{cy_side_channel}.
While the class of attacks can be broadened, for illustrative examples let us consider the following adversarial capabilities:

\begin{enumerate}[\it i\normalfont )]

\item \textit{Integrity violation}: 
An adversary can get a foothold in the system \cite{cy_side_channel,taskshuffler}. For example, an adversary may insert a malicious task that respects the real-time guarantees of the system to avoid immediate detection, and/or compromise one or more existing real-time tasks. 
Such a task can be used to manipulate sensor inputs and actuator commands for instance and/or modify system behavior in undesirable ways.


\item \textit{Denial of Service (DoS)}: The attacker may take control of the real-time task(s) and perform \textit{system-level} resource (\eg CPU, disk, memory, \etc) exhaustion.  In 
particular, when critical tasks are scheduled to run
an advanced attacker may capture I/O or network ports and perform \textit{network-level} attacks to tamper with the confidentiality and integrity (\viz safety) of the system.

\end{enumerate}

Threats to communications are usually dealt with by integrating cryptographic protection mechanisms. From an RTS perspective this increases the execution time of existing real-time tasks~\cite{xie2007improving, lin2009static}. 
\coolname is different from earlier work  in which integration of security impacted the schedulability \cite{sibin_RT_security_journal,sg1, sg2}, required modification of the existing schedulers \cite{xie2007improving, lin2009static}, or necessitated architectural modifications \cite{slack_cornell,securecore}. In this work, we focus on incorporating security mechanisms into legacy systems in which added security tasks are \textit{not} allowed to violate the temporal requirements (in either the \pve or \ave modes) and must have \textit{minimal} impact on the schedule of existing real-time tasks.

Let us consider an RTS (say an avionics electronic control unit) developed using a multi-vendor model \cite{sg2}, \viz  its components are manufactured and integrated by different vendors. For example, tasks in the system component manufactured by vendor $v_i$ are very sensitive and considered classified or mission-critical (\eg images captured by the camera on the surveillance UAV). It may be undesirable for any vendor $v_j \neq v_i$ to gain unintended information about sensitive contents, even if, say, vendor $v_j$ is trusted with control tasks for controlling the RTS. Similarly, the control laws from vendor $v_j$ may contain a proprietary algorithm and vendor $v_j$ may not want other vendors to gain knowledge about the algorithm. Protected communications and network monitoring/detection mechanisms are necessary but insufficient to deal with such threats. Therefore, \textit{additional security tasks} may need to be added into the system to deal with such threats \cite{sibin_deeply}. 

The security mechanisms could be protection, detection or response mechanisms, depending on the system requirements. For example, a sensor measurement correlation task may be added to detect sensor manipulation, a change detection task may be added to detect intrusions or additional state-cleansing tasks \cite{sg1, sg2, sibin_RT_security_journal} can be added to deal with stealthy adversaries trying to glean sensitive information through side channels.

It is worth mentioning that the addition of such security mechanisms may necessitate changes to the schedule of real-time tasks \cite{xie2007improving, lin2009static, sg1, sg2, sibin_RT_security_journal}. In contrast, \coolname aims to integrate such security tasks \textit{without} impacting the timeliness constraints (\ie schedulability) required for safe operation (in both modes) and retaining the original schedule of real-time tasks most of the time (\eg in \pve mode when security tasks are executing opportunistically with lowest priority). We highlight that rather than designing specific intrusion detection tasks that target specific attack behaviors, the generic framework proposed in this work allows one to integrate a given security mechanism (referred to as \textit{security tasks}) into the system without perturbing the system parameters (\eg period of the real-time tasks, execution order, \etc).

\subsection{Overview of \coolname} \label{subsec:timeshield_overview}


   \begin{figure}[!t]
\includegraphics[width=\linewidth]{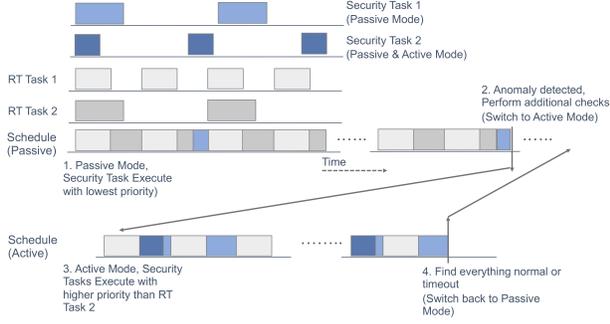}
\caption{\coolname: Flow of operations depicting the \pve and \ave modes for the security tasks.}
\label{fig:flow_operation}
 \end{figure}

As illustrated in Fig.~\ref{fig:flow_operation}, \coolname improves the security posture of the system (that contains a set of real-time tasks) by integrating additional security tasks and allowing them to execute in two different \textit{modes} (\viz \pve and \ave). 
If the system is deemed to be clean (\ie not compromised), security routines can execute \textit{opportunistically}\footnote{Which is also the default mode of operation.} (\eg when other real-time tasks are not running). However if any anomaly or unusual behavior is suspected, the security policy may switch to \ave mode (\eg more fine-grained checking or response) and execute with \textit{higher priority} for a \textit{limited amount of time} (since our goal is to ensure security with minimum perturbation of the scheduling order of the real-time tasks). 
The security routines may go back to normal (\eg~\pve) mode  if: 
\begin{itemize}
\item No anomalous activity is found within a predefined time duration, say $T^{AC}$; or
\item The intrusion is detected and malicious entities are removed (or an alarm triggered if human intervention is required). 
\end{itemize}

Although we allow the security tasks to execute with higher priority than some of the real-time tasks in \ave mode, the proposed framework ensures that the timeliness constraints (\eg deadlines) for \textit{all} of the real-time tasks are always satisfied in \textit{both} modes. By using this strategy, \coolname not only enables \textit{compatibility with legacy systems} (\eg in normal situation real-time scheduling order is not perturbed), but also provides \textit{flexibility to promptly deal with anomalous behaviors} (\ie the security tasks are promoted to higher priority so that they can experience less preemption and achieve better response times).

\subsection{System Model} \label{subsec:sec_task}

\subsubsection{Real-Time Tasks}
In this paper we consider the widely used fixed-priority sporadic task 
model \cite{sporadic_task}. Let us consider a uniprocessor system consisting of $m$ fixed-priority
sporadic real-time tasks $\Gamma_R = \lbrace \tau_1, \tau_2, \cdots , \tau_m \rbrace$. Each real-time task $\tau_j \in \Gamma_R$ is characterized by $(C_j, T_j , D_j)$, where $C_j$ is the WCET, $T_j$ is the minimum inter-arrival time (or period) between successive releases and $D_j$ is the relative deadline.  We assume that priorities are distinct and assigned according to the rate monotonic (RM) \cite{Liu_n_Layland1973} order. 



The processor utilization of $\tau_j$ is defined as $U_j = \frac{C_j}{T_j}$. Let $hp_R(\tau_j)$ and $lp_R(\tau_j)$ denote the sets of real-time tasks that have higher and lower priority than $\tau_j$, respectively. We assume that the real-time task-set $\Gamma_R$ is \textit{schedulable} by a fixed-priority preemptive scheduling algorithm. Therefore, the worst-case response time 
$w_i$ is less than or equal to the deadline $D_i$ and the following inequality is satisfied for all tasks $\tau_j \in \Gamma_R$:
$w_j \leq D_j$,
where $w_j = w_j^{k+1} = w_j^k$ is obtained by the following recurrence relation \cite{res_time_rts}:
\begin{equation} \label{eq:wcrt_recurrence}
\begin{aligned}
w_j^0 = C_j, \quad 
w_j^{k+1} = C_j + \sum_{\tau_h \in hp_R(\tau_j)} \left\lceil \frac{w_j^k}{T_h} \right\rceil C_h.
\end{aligned}
\end{equation}
In Eq.~(\ref{eq:wcrt_recurrence}), $\sum\limits_{\tau_h \in hp_R(\tau_j)} \left\lceil \frac{w_j^k}{T_h} \right\rceil C_h$ is the worst-case interference to  $\tau_j$ due to preemption by the tasks with higher priority than $\tau_j$ (\eg $hp_R(\tau_j)$). 
The recurrence will have a solution if $w_j^{k+1} = w_j^k$ for some $k$.  

\subsubsection{Security Tasks}



With a view of integrating  security into the system, let us add additional fixed-priority security tasks that will be executed in \pve and \ave modes. We model \pve and \ave mode security tasks as independent \textit{sporadic tasks}. The \pve and \ave mode tasks are denoted by the sets $\Gamma_S^{pa} = \lbrace \tau_1, \tau_2, \cdots , \tau_{n_p} \rbrace$ and $\Gamma_S^{ac} = \lbrace \tau_1, \tau_2, \cdots , \tau_{n_a} \rbrace$, respectively.
We assume that security tasks in both modes follow RM priority order. Each security task $\tau_i \in \lbrace \Gamma_S^{pa} \cup \Gamma_S^{ac} \rbrace$ is characterized by the tuple $(C_i, T_i^{des}, T_i^{max}, \omega_i)$, where $C_i$ is the WCET, $T_i^{des}$ is the most desired period between successive releases (hence $F_i^{des} = \frac{1}{T_i^{des}}$ is the desired execution frequency of a security routine) and $T_i^{max}$ is the maximum allowable period beyond which security checking by $\tau_i$ may not be effective. The parameter $\omega_i > 0$ is a designer-provided weighting factor that may reflect the criticality of the security task\footnote{As an example, the default configuration of Tripwire \cite{tripwire}, an intrusion detection system (IDS) for Linux that we use as case study in Section \ref{subsec:implementation_BBB}, has different criticality levels (\viz weights), \ie  \textit{High} (for scanning files that are significant points of vulnerability), \textit{Medium} (for non-critical files that are of significant security impact) and so forth.} $\tau_i$. 
Critical security tasks would have larger $\omega_i$. The security tasks have implicit deadlines, \eg $D_i = T_i, \forall \tau_i$ that implies security tasks should complete before 
their next monitoring instance. 
We do not make any specific assumptions about the security tasks in different modes. For instance, both \pve and \ave mode task-sets may contain completely different sets of tasks (\eg $\lbrace \Gamma_S^{pa} \cap \Gamma_S^{ac} \rbrace = \emptyset$) or may contain (partially) identical tasks with different parameters (\eg period and/or criticality requirements). 

In \pve mode, security tasks are executed with \textit{lower priority} than the real-time tasks. Hence the security tasks do not have any impact on real-time tasks and cannot perturb the real-time scheduling order. In \ave mode, we allow the security tasks to execute with a priority higher than that of certain low priority real-time tasks. This provides us with a trade-off mechanism between security (\eg responsiveness) and system constraints (\eg scheduling order of real-time tasks). Since the task priorities are distinct, there are $m$ priority-levels for real-time tasks (indexed from $0$ to $m-1$ where level $0$ is the highest priority). Among the $m$ priority-levels, we assume that \ave mode security tasks can execute with a priority-level up to $l_S$ $(0 < l_S \leq m),~ l_S \in \mathbb{Z}$. 

Although any period $T_i$ within the range $T_i^{des} \leq T_i \leq T_i^{max}$ is acceptable for \pve (\eg $\tau_i \in \Gamma_S^{pa}$) and \ave (\eg $\tau_i \in \Gamma_S^{ac}$) mode  security tasks, the actual period $T_i$ is not known a priori. Furthermore, for \ave mode security tasks (\eg $\tau_i \in \Gamma_S^{ac}$), we need to find out the suitable priority level $l \in [l_S, m]$. Therefore our goal is to find the \textit{suitable period} (for both \pve and \ave mode security tasks) as well as the \textit{priority-level} (for \ave mode security tasks) that achieve the best trade-off between schedulability and defense against security breaches without violating the real-time constraints.

\section{Period Adaptation}
\label{sec:period_adapt}

As already mentioned, one fundamental problem in integrating security tasks is to determine \textit{which} security tasks will be running \textit{when}.
This brings up the challenge of determining the \textit{right periods} (\viz the minimum inter-execution times) for the security tasks. For instance, some critical security routines may be required to execute more frequently than others. However, if the period is too short (\eg the security task repeats too often) then it will use too much of the processor time and eventually lower the overall system utilization. As a result, the security mechanism itself might prove to be a hindrance to the system and reduce the overall functionality or, worse, safety. In contrast, if the period is too long, the security task may not always detect violations, since attacks could be launched between two instances of the security task.

One may wonder why we cannot assign the desired period (\eg $T_i = T_i^{des}$) in both \pve and \ave modes and set the \ave mode priority level  as $l = l_S$ so that  the security tasks can always execute with the desired frequency (\ie $F_i^{des} = \frac{1}{T_i^{des}}$) and experience less interference (\eg preemption) from real-time tasks. However, since our goal is to integrate security mechanisms in legacy systems with minimal\footnote{In \ave mode \coolname does not introduce any timing violations for the real-time tasks, but their execution might
be delayed due to interference from high-priority security tasks (\eg the tasks with priority-level $l \in [l_S, m]$).} or no perturbation, setting $T_i = T_i^{des}, ~\forall \tau_i$ in either or both mode(s) may significantly perturb the real-time scheduling order. If the schedulability of the system is not analyzed after the perturbation, some (or all) of the real-time tasks may miss their deadlines and thus the main safety requirements of the system will be threatened. The same argument is also true for \ave mode if we set $l = l_S$ (or arbitrarily from the range $[l_S, m]$)  and do not perform schedulability analysis carefully.

\subsubsection*{Tightness of the Monitoring}

As mentioned earlier, the actual period as well as the priority-levels of the security tasks  are unknown and we need to \textit{adapt} the periods within acceptable ranges. We measure the security of the system by means of  \textit{achievable periodic monitoring}. Let $T_i$ be the period of the security task $\tau_i \in \lbrace  \Gamma_S^{pa} \cup \Gamma_S^{ac} \rbrace$ that needs to be determined. Our goal is to minimize the gap between the achievable period $T_i$ and the desired period $T_i^{des}$ and therefore we define the following metric:
\begin{equation}
\eta_i = \frac{T_i^{des}}{T_i},
\end{equation}
that denotes the \textit{tightness} of the frequency of periodic monitoring for the security task $\tau_i$. Thus $\eta^{pa} =  \sum\limits_{\tau_i \in \Gamma_S^{pa}} \omega_i \eta_i$ and $\eta^{ac} =  \sum\limits_{\tau_i \in \Gamma_S^{ac}} \omega_i \eta_i$ denote the \textit{cumulative tightness} of the achievable periodic monitoring for \pve and \ave mode, respectively.
This monitoring frequency metric, provides for instance,  one way to trade-off security with schedulability. Recall that
if the interval between consecutive monitoring events is too large, the adversary may remain undetected and harm the system between two invocations of the security task. Again, a very frequent execution of security tasks may impact the schedulability of the real-time tasks. This metric $\eta^{(\cdot)}$ will allow us to execute the security routines with a frequency closer to the desired one while respecting the temporal constraints of the other real-time tasks.

\subsection{Problem Overview} \label{subsec:proble_overview}

One may wonder why we cannot schedule the security tasks in the same way that the existing real-time tasks are scheduled. For instance, a simple approach to integrating security tasks in \pve mode without perturbing real-time scheduling order is to execute security tasks at a \textit{lower priority} than all real-time tasks. Hence, the security routines will be executing only during slack times when no other higher-priority real-time tasks are running. Likewise, in \ave mode, security tasks can be executed at a lower priority than more critical, high-priority real-time tasks. 
Hence, the security tasks will only be executing when other real-time tasks with priority-levels higher than $l_S$ are not running. 

When both real-time and security tasks follow RM priority order, we can formulate a nonlinear optimization problem for \pve mode with the following constraints that maximizes the cumulative tightness of the frequency of periodic monitoring:

\begin{myoptimizationproblem} \label{opt:period_adapt_ns_pve}
\vspace*{-2.0em}
\begin{subequations}
\begin{align}
 & \hspace*{5em} \underset{\mathbf{T}^{\boldsymbol {pa}}}{\operatorname{max}}~~\eta^{pa}  \nonumber \\
\hspace*{-0.45em}\text{Subject to:}~~ \nonumber \\
\sum_{\tau_i \in \Gamma_S^{pa}}\frac{C_i}{T_i} &\leq (m+n_p)(2^{\frac{1}{m+n_p}} - 1) - \sum_{\tau_j \in \Gamma_R}\frac{C_j}{T_j} \label{eq:period_ns_const_2_pve}\\
T_i &\geq \underset{\tau_j \in \Gamma_{R} }{\operatorname{max}} T_j \quad \quad \forall \tau_i \in \Gamma_S^{pa} \label{eq:period_ns_const_rm_pve} \\
T_i^{des} &\leq T_i \leq T_i^{max}  \quad \forall \tau_i \in \Gamma_S^{pa} \label{eq:period_const_1_pve}
\end{align}
\end{subequations}
\end{myoptimizationproblem}  
\hspace{-1.9em}
where $\mathbf{T}^{\boldsymbol{pa}} = [ T_1, T_2, \cdots, T_{n_p} ]^{\mathsf{T}}$ is the optimization variable for \pve mode that needs to be determined. The constraint in Eq.~(\ref{eq:period_ns_const_2_pve}) ensures that the utilization 
of the security tasks are within the remaining RM utilization bound \cite{Liu_n_Layland1973}. The RM priority order for real-time and security tasks is ensured by the constraints in Eq.~(\ref{eq:period_ns_const_rm_pve}), while  Eq.~(\ref{eq:period_const_1_pve}) ensures the restrictions on periodic monitoring. 

Recall that in \ave mode,  we allow the security tasks to execute when the real-time tasks with priority-levels higher than $l_S$ are not running. Hence, to ensure the RM priority order in \ave mode, we need to modify the constraints in Eq.~(\ref{eq:period_ns_const_rm_pve}) as follows:
\begin{equation}
T_i \geq \underset{\tau_j \in \Gamma_{R_{hp(l_S)}} }{\operatorname{max}} T_j, \quad \forall \tau_i \in \Gamma_S^{ac} \label{eq:period_ns_const_rm_ave}
\end{equation}
where $ \Gamma_{R_{hp(l_S)}}$ represents the set of real-time tasks that are higher priority than level $l_S$. In addition, 
the constraints in Eq.~(\ref{eq:period_ns_const_2_pve}) and 
Eq.~(\ref{eq:period_const_1_pve}) also need to be updated to consider \ave mode task-sets (\eg $\Gamma_S^{ac}$) and the number of active mode security tasks ($n_a$). Thus for \ave mode we can formulate an optimization problem similar to that of $\mathbf{P}\mathbf{\ref{opt:period_adapt_ns_pve}}$ with the objective function: $\underset{\mathbf{T}^{\boldsymbol{ac}}}{\operatorname{max}} ~~  \eta^{ac}$, where $\mathbf{T}^{\boldsymbol{ac}} = [ T_1, T_2, \cdots, T_{n_a} ]^{\mathsf{T}}$ is the \ave mode optimization variable.

One of the limitations of the above approach is that the overall system utilization is limited by the RM bound which has the theoretical upper bound of processor utilization only about $\lim\limits_{n \rightarrow \infty} n(2^{\frac{1}{n}}-1) = \ln 2 \approx 69.31\%$ \cite{Liu_n_Layland1973}, where $n$ is the total number of tasks under consideration. Further, the security tasks' periods need to satisfy the constraints in Eq.~(\ref{eq:period_ns_const_rm_pve}) and Eq.~(\ref{eq:period_ns_const_rm_ave}) (for \pve and \ave modes, respectively) to follow RM priority order. In addition, instead of focusing only on optimizing the periods of the security tasks, \coolname aims to provide a \textit{unified} framework that can achieve other security aspects (\viz responsiveness). Thus we follow an alternative approach similar to one we proposed in earlier work\footnote{The approach we proposed in our earlier work \cite{mhasan_rtss16} is analogous to the \pve mode of \coolname.} \cite{mhasan_rtss16}. Specifically, we had proposed 
to use a \textit{server} \cite{server_ab_uk} to execute security tasks. Our security server is motivated by the needs of hierarchical scheduling \cite{periodic_server_qp}. Under hierarchical
scheduling, the system is composed of a set of
components (\eg real-time tasks and a security server, in our context) and each of which comprises multiple tasks or subcomponents (\eg security tasks). The server abstraction not only allows us to provide better isolation between real-time and security tasks, but also enables us to integrate additional security properties (such as responsiveness) as we discuss in the following. 


\section{The Security Server} \label{sec:sec_server}


The server \cite{server_ab_uk} is an abstraction that provides execution time to the
security tasks according to a predefined scheduling algorithm.  Our proposed security server is characterized by the
\textit{capacity} $Q$ and \textit{replenishment period} $P$ and works as
follows. 
The server is executed with lowest-priority in \pve mode. However, in 
\ave mode, the server can switch to any allowable priority-level\footnote{Calculation of the server priority-level is described in Section \ref{sec:algo}.} within the range $[l_S, m]$. If any security
task is activated at time $t$, 
then the server
is activated with capacity $Q$ and the next
replenishment time is set as $t + P$.  
When the server is
 scheduled, it executes the security tasks according to its own scheduling
policy. In this work we assume that the server schedules the security tasks using fixed-priority RM scheduling. When a
security task executes, the current available capacity is decremented
accordingly. The server can be preempted by the scheduler to service 
real-time tasks. When the server is preempted, the currently available capacity
is not decremented. If the available capacity becomes zero and some
security task has not yet finished, then the server is suspended until its next
replenishment time ($t'$). At time $t'$, the
server is recharged to its full capacity $Q$, the next replenishment time is
set as $t' + P$, and the server is executed again. When the last security task
has finished executing and there is no other pending task in the server, the
server will be suspended. Also, the server will become inactive if there are no
security tasks ready to execute. 



\subsection{Reformulation of the Period Adaptation Problem using Servers} \label{subsec:period_adapt_server}

When security tasks execute within the server, we need to modify the constraints in the period adaption problem considering the server parameters $Q$ and $P$. In the following we briefly discuss how to customize the period adaptation problem with the inclusion of the server.


 Let us use $UB_{\mathcal{S}(Q, P), \Gamma}$ to denote the utilization bound for the set of tasks $\Gamma$ executing within the server.  
When the smallest period of the task is greater than or equal to $3P - 2Q$, it has been shown \cite{serverBound_raj} that the upper bound of the utilization factor for the security tasks is given by
$
UB_{\mathcal{S}(Q, P), \Gamma} = n \left[ \left( \tfrac{3 - \tfrac{Q}{P}}{3 - 2 \tfrac{Q}{P}} \right)^{\frac{1}{n}} - 1 \right]$,
where $n$ is number of tasks in the set $\Gamma$.

Thus with the inclusion of the server in \pve mode, we can modify the constraints in Eqs. (\ref{eq:period_ns_const_2_pve}) and (\ref{eq:period_ns_const_rm_pve}) as follows:
\begin{subequations}
\begin{align} 
\sum_{\tau_i \in \Gamma_S^{pa}}\frac{C_i}{T_i} &\leq  n_p \left[ \left( \tfrac{3 - \tfrac{Q^{pa}}{P^{pa}}}{3 - 2 \tfrac{Q^{pa}}{P^{pa}}} \right)^{\frac{1}{n_p}} - 1 \right] \label{eq:period_const_2_pve} \\ 
T_i &\geq 3P^{pa} - 2Q^{pa}, \quad \forall \tau_i \in \Gamma_S^{pa}. \label{eq:period_const_4_pve}
\end{align}
\end{subequations}
Therefore, selection of the periods for security tasks in \pve mode 
is a nonlinear constrained optimization problem that can be formulated as follows: 

\begin{myoptimizationproblem} \label{opt:period_adapt_server_pve}
\vspace*{-2.0em}
\begin{subequations}
\begin{align}
\quad \quad \underset{\mathbf{T}^{\boldsymbol{pa} }}{\operatorname{max}} \sum_{\tau_i \in \Gamma_S^{pa}} \omega_i \frac{T_i^{des}}{T_i}, ~~ 
\text{Subject to:~  (\ref{eq:period_const_2_pve}), (\ref{eq:period_const_4_pve}),  (\ref{eq:period_const_1_pve})} \nonumber.
\end{align}
\end{subequations}
\end{myoptimizationproblem}
\hspace{-1.9em}
where $Q^{pa}$ and $P^{pa}$ are the server capacity and replenishment period in \pve mode, respectively. The formulation of the \pve mode period adaptation problem presented above is similar to that we proposed in earlier work \cite{mhasan_rtss16}.

Similarly, in \ave mode, the period adaptation problem can be reformulated as follows:

\begin{myoptimizationproblem} \label{opt:period_adapt_server_ave}
\vspace*{-2.0em}
\begin{subequations}
\begin{align}
\underset{\mathbf{T}^{\boldsymbol{ac}}}{\operatorname{max}} & \sum_{\tau_i \in \Gamma_S^{ac}} \omega_i \frac{T_i^{des}}{T_i} ~~ \quad  \nonumber \\
\text{Subject to:~} 
\sum_{\tau_i \in \Gamma_S^{ac}}\frac{C_i}{T_i} &\leq  n_a \left[ \left( \tfrac{3 - \tfrac{Q^{ac}}{P^{ac}}}{3 - 2 \tfrac{Q^{ac}}{P^{ac}}} \right)^{\frac{1}{n_a}} - 1 \right] \label{eq:period_const_2_ave} \\
T_i &\geq 3P^{ac} - 2Q^{ac} \quad \forall \tau_i \in \Gamma_S^{ac} \label{eq:period_const_4_ave} \\
T_i^{des} &\leq T_i \leq T_i^{max}  \quad ~~\forall \tau_i \in \Gamma_S^{ac} \label{eq:period_const_1_ave}
\end{align}
\end{subequations}
\end{myoptimizationproblem}
\hspace{-1.9em}
where $Q^{ac}$ and $P^{ac}$ are the server capacity and replenishment period in \ave mode, respectively.

\subsection{Selection of the Server Parameters} \label{sec:server_param_selec}

The period adaptation problem illustrated in Section \ref{subsec:period_adapt_server}
  is derived based on a given set of server parameters, \eg $(Q^{(\cdot)}, P^{(\cdot)})$. However, a fundamental problem is to find a suitable pair of server capacity $Q^{(\cdot)}$ and replenishment period $P^{(\cdot)}$ that respects the real-time constraints of the tasks in the system. Our approach to selecting the server parameters in \pve and \ave mode is described below.

\subsubsection{Parameter Selection in Passive Mode} \label{subsec:param_pve}

Recall that in \pve mode, the server will execute with the lowest priority to have compatibility with existing real-time tasks. Since the security tasks execute within the server, we need to ensure the following two constraints:
\begin{itemize}

\item \textit{The server is schedulable}: that is  the server's capacity and interference from higher priority real-time tasks are less than the replenishment period; and

\item \textit{The security tasks are schedulable}: the minimum \textit{supply} by the server to the security tasks  is greater than the worst-case workload generated by the security tasks.

\end{itemize}

Note that since the server is running with lowest priority, the real-time constraints 
(\eg
$w_j \leq D_j, \forall \tau_j \in \Gamma_R$)
and the task execution order are not affected in the \pve mode.  Based on the above two constraints, we illustrate an approach for determining the server parameters by formulating it as a \textit{constraint optimization problem}.

The security server is referred to as \textit{schedulable} if the worst-case response time of the server does not exceed its replenishment period \cite{server_ab_uk}. Thus, following an approach similar to ones in  earlier work \cite{mhasan_rtss16, mn_gp},  the \textit{server schedulability constraint} can be represented as follows: 
 \begin{equation} \label{eq:ser_con1_pve}
 Q^{pa} + \Delta_{\mathcal{S}^{pa}} \leq P^{pa}
 \end{equation}
 where $\Delta_{\mathcal{S}^{pa}} = \sum\limits_{\tau_h \in hp_R(\tau_{\mathcal{S}^{pa}})} \left( \frac{P^{pa}}{T_h} + 1 \right)  C_h$ is the worst-case interference experienced by the server when preempted by the higher priority real-time tasks.
In the above equation, the set of real-time tasks with higher priority than the server (\ie $hp_R(\tau_\mathcal{S}^{pa}) = \Gamma_R$) is fixed.

Let us use $hp_S^{pa}(\tau_i)$ to denote the set of \pve mode security tasks that are higher priority than $\tau_i \in \Gamma_S^{pa}$. To ensure schedulability of the security tasks, we can derive the \textit{minimum supply} of the server delivered to the security tasks by using the periodic resource model from the literature  \cite{periodic_server_qp, mn_gp, mhasan_rtss16}. In particular, the constraints on the server supply to ensure \textit{schedulability of the security tasks} \cite{mhasan_rtss16} can be expressed as:
\begin{equation} \label{eq:ser_con2_pve}
 \frac{Q^{pa}}{P^{pa}} \left[ T_i - (P^{pa} - Q^{pa}) - \Delta_{\mathcal{S}^{pa}}  \right] \geq I_i^{pa}, ~~\forall \tau_i \in \Gamma_S^{pa} 
\end{equation}
where $I_i^{pa} = C_i + \sum\limits_{\tau_h \in hp_S^{pa}(\tau_i)} \left\lceil \frac{T_i}{T_h} \right\rceil C_h$ is the worst-case workload generated by the security task $\tau_i$ and $hp_S^{pa}(\tau_i)$ during the time interval of $T_i$. This workload is a constant for a given input. 

Since we need to ensure maximal processor utilization for the security tasks without violating the real-time constraints of the system, we define the following objective function:
$\underset{Q^{pa}, P^{pa}}{\operatorname{max}}~ \frac{Q^{pa}}{P^{pa}}$. 
With 
this objective function
and the constraints in Eqs.~(\ref{eq:ser_con1_pve})-(\ref{eq:ser_con2_pve}), the \pve mode server parameter selection problem can be formulated as follows:

\begin{myoptimizationproblem} \label{opt:server_param_pve}
\vspace*{-2.0em}
\begin{subequations}
\begin{align}
\hspace*{5em}
\underset{Q^{pa}, P^{pa}}{\operatorname{max}}~ \frac{Q^{pa}}{P^{pa}}, \quad  
\text{Subject to:~  (\ref{eq:ser_con1_pve}), (\ref{eq:ser_con2_pve})} \nonumber
\end{align}
\end{subequations}
\end{myoptimizationproblem}
\hspace{-1.9em} 
where server parameters $Q^{pa}$ and $P^{pa}$ are the optimization variables.

\subsubsection{Parameter Selection in Active Mode} \label{subsec:param_ave}



In \ave mode, the security server is \textit{no longer the lowest priority task}. Since the server can execute with priority $l_S$, there could be up to $m - l_S$ low priority real-time tasks than that of the server. Thus we need to ensure the schedulability of the real-time tasks that are executing with a priority lower than the server. Hence, in addition to the constraints described in Section \ref{subsec:param_pve} (\ie Eqs.~(\ref{eq:ser_con1_pve})-(\ref{eq:ser_con2_pve})), we need to consider the following:

\begin{itemize}
\item \textit{The real-time tasks with lower priority than the server are schedulable}: that is, the interferences from the server and other higher priority real-time tasks do not violate the deadlines for these low-priority tasks.
\end{itemize}

We therefore define the following constraints to ensure the \textit{schedulability of the low-priority real-time tasks}:
\begin{equation}\label{eq:ser_con3_ave}
\begin{aligned}
C_j + \hspace*{-1em} \sum_{\tau_h \in hp_R(\tau_j)} \left\lceil \frac{D_j}{T_h} \right\rceil C_h + \left( \frac{D_j}{P^{ac}} + 1 \right) Q^{ac} \leq D_j,~~ \\  
\forall \tau_j \in lp_R(\tau_\mathcal{S}^{ac}) 
\end{aligned}
\end{equation}
where $\sum\limits_{\tau_h \in hp_R(\tau_j)} \left\lceil \frac{D_j}{T_h} \right\rceil C_h$ is the interference experienced by $\tau_j$ from other real-time tasks and $\left( \frac{D_j}{P^{ac}} + 1 \right) Q^{ac}$ is the worst-case interference caused to $\tau_j$ by the server in \ave mode. As illustrated in Section \ref{sec:algo}, we iterate through the allowable priority ranges (\eg $[l_S, m]$) to find the server priority in \ave mode. Note that for a given priority-level, the set of tasks $lp(\tau_\mathcal{S}^{ac})$ is predefined. Thus the only variables for the constraints in Eq.~(\ref{eq:ser_con3_ave}) are the server capacity $Q^{ac}$ and replenishment period $P^{ac}$.

Let us use $hp_S^{ac}(\tau_i)$ to denote the set of \ave mode security tasks that are higher priority than $\tau_i \in \Gamma_S^{ac}$. Just as in $\mathbf{P}\mathbf{\ref{opt:server_param_pve}}$ we can now formulate the \ave mode parameter selection problem as follows:

\begin{myoptimizationproblem} \label{opt:server_param_ave}
\vspace*{-2.0em}
\begin{subequations}
\begin{align}
\underset{Q^{ac}, P^{ac}}{\operatorname{max}}~ \frac{Q^{ac}}{P^{ac}}, ~~& 
\text{Subject to:   (\ref{eq:ser_con3_ave})~and} 
\nonumber \\
Q^{ac} + \hspace*{-1em} \sum\limits_{\tau_h \in hp_R(\tau_{\mathcal{S}^{ac}})} \left( \frac{P^{ac}}{T_h} + 1 \right)  C_h &\leq P^{ac} \label{eq:ser_con1_ave}
  \\
   \frac{Q^{ac}}{P^{ac}} \left[ T_i - (P^{ac} - Q^{ac}) - \Delta_{\mathcal{S}^{ac}}  \right] &\geq I_i^{ac} ~~ \forall \tau_i \in \Gamma_S^{ac} \label{eq:ser_con2_ave}
\end{align}
\end{subequations}
\end{myoptimizationproblem}
\hspace{-1.9em} 
where the set of real-time tasks with higher priority than the server (\ie $hp_R(\tau_\mathcal{S}^{ac}) \subset \Gamma_R$) is a constant for a given priority-level and $I_i^{ac} = C_i + \sum\limits_{\tau_h \in hp_S^{ac}(\tau_i)} \left\lceil \frac{T_i}{T_h} \right\rceil C_h$ is the worst-case workload generated by the security task $\tau_i$ and $hp_S^{ac}(\tau_i)$. Note that the schedulability of the higher priority real-time tasks (\eg $\forall \tau_j \in hp_R(\tau_\mathcal{S}^{ac})$) is already ensured by definition. 

\begin{remark}
The formulation of the period adaptation and server parameter selection problems are nonlinear constraint optimization problems and are nontrivial to solve in their current form. However, these problems can be transformed  into a geometric programming (GP) \cite{GP_tutorial} problem. 
In addition, it is also possible to reformulate the non-convex GP representation into equivalent convex form that can be solved using known algorithms such as \textit{interior point} \cite[Ch. 11]{boyd_book} method. For details of this reformulation, we refer the readers to Appendix. 
\end{remark}



\subsection{Discussion on Mode Switching} \label{subsec:mode_switch_diss}

As mentioned earlier, by default, \coolname operates in \pve mode. However, when a malicious activity is suspected, a \pve-to-\ave mode change request will be issued. Similarly,  an \ave-to-\pve mode change request will be placed if the system seems clean after fine-grained checking, or a malicious entity is found and removed. 
 In steady-state (\eg when security tasks are executing in \pve or \ave mode), the schedulability of the real-time tasks is already guaranteed by the analysis presented in Section~\ref{sec:server_param_selec}. 
 
 
When \coolname switches from \pve mode to \ave mode, the schedulability of real-time tasks will not be affected. The reason this that \textit{all} the real-time tasks are higher priority than the security tasks in \pve mode and hence do not suffer any additional interference from security tasks during mode change. Therefore, the schedulability of real-time tasks during \pve-to-\ave mode switching is already covered by steady-state analysis (Section \ref{subsec:param_pve}).
 

During \ave-to-\pve mode switching, observe that schedulability of the real-time tasks that have a priority higher than the  server (\ie $hp_R(\tau_\mathcal{S}^{ac})$) is not affected. When the mode switch request is issued, the \ave mode server (and the security tasks) stop execution and the control is then switched to the lowest priority \pve mode server.
Note that the constraints in Eq.~(\ref{eq:ser_con3_ave}) that ensures the schedulability of the low-priority real-time tasks already captures the worst-case interference introduced by the server. Hence the server will not impose any more interference (even if the mode switch is performed in the middle of the execution of a
busy interval) on the low-priority real-time tasks than what we have calculated in the steady-state analysis (Section \ref{subsec:param_ave}). Therefore if both the \pve and \ave modes task-sets are schedulable, the system will also be schedulable with mode changes.

\section{Algorithm Development} \label{sec:algo}

We develop a simple scheme to obtain the security task's period (for both \pve and \ave mode) and priority-level (for \ave mode). The overall algorithm, Algorithm \ref{alg:sec_schd}, works as follows.

	\renewcommand{\algorithmicforall}{\textbf{for each}}
    \renewcommand\algorithmiccomment[1]{%
 {\it /* {#1} */} %
}
\renewcommand{\algorithmicrequire}{\textbf{Input:}}
    \renewcommand{\algorithmicensure}{\textbf{Output:}}
    
   \newcommand{\Function}[1]{\textbf{function}~\textsc{#1}}
   \newcommand{\EndFunction}{\textbf{end function}}

		\begin{algorithm}[!ht]
        \algsetup{linenosize=\relsize{-0.9}\footnotesize}
  \relsize{-0.9}\footnotesize
			\begin{algorithmic}[1]
				\REQUIRE Set of real-time tasks, $\Gamma_R$, \pve and 
               \ave mode security tasks $\Gamma_S^{pa}$ and $\Gamma_S^{ac}$, allowable priority ranges $[l_S,m]$
    \ENSURE The tuple $\left\lbrace l^*, \mathbf{T}^{\boldsymbol{pa}}, Q^{pa}, P^{pa},  \mathbf{T}^{\boldsymbol{ac}}, Q^{ac}, P^{ac} \right\rbrace$, \eg \ave mode server priority-level,  \ave and \pve mode periods of the security tasks and \ave and \pve mode server parameters if the task-set is schedulable; $\mathsf{Unschedulable}$	 otherwise	
					\vspace{0.4em}

                \STATE Obtain \pve and \ave mode parameters using the functions {\sc {PassiveModeParamSelection($\Gamma_R$, $\Gamma_S^{pa}$)}} and {\sc {ActiveModeParamSelection($\Gamma_R$, $\Gamma_S^{ac}$, $l_S$)}}
             
              \IF {$\mathsf{Solution~ Found ~in~ BOTH ~Modes}$}
                    \STATE  \COMMENT{return the parameters}
                    \STATE \textbf{return} $\left\lbrace l^*, \mathbf{T}^{\boldsymbol{pa}}, Q^{pa}, P^{pa},  \mathbf{T}^{\boldsymbol{ac}}, Q^{ac}, P^{ac} \right\rbrace$
                     
                    \ELSE
\STATE \COMMENT{not possible to integrate security tasks in the system}	                   \STATE \textbf{return} $\mathsf{Unschedulable}$  
                    \ENDIF
         \\\hrulefill           
                \vspace*{0.5em}
              \STATE \Function{PassiveModeParamSelection($\Gamma_R$, $\Gamma_S^{pa}$)}
              
               \begin{ALC@g}
        \STATE Initialize \pve mode period $T_i := T_i^{des}, \forall \tau_i \in \Gamma_S^{pa}$ 
        	\STATE Solve $\mathbf{P}\mathbf{\ref{opt:server_param_pve}}$ to obtain server parameters
					\IF {$\mathsf{Solution Found}$}
                    \STATE Solve $\mathbf{P}\mathbf{\ref{opt:period_adapt_server_pve}}$ to obtain security periods
                    \IF {$\mathsf{Solution Found}$}
                    \STATE  \COMMENT{return the parameters}
                    \STATE \textbf{return} $\mathbf{T}^{\boldsymbol{pa}}$, $Q^{pa}$, $P^{pa}$ where $Q^{pa}$, $P^{pa}$ and $\mathbf{T}^{\boldsymbol{pa}}$  are the solutions obtained by $\mathbf{P}\mathbf{\ref{opt:server_param_pve}}$ and $\mathbf{P}\mathbf{\ref{opt:period_adapt_server_pve}}$
                     
                    \ENDIF
                    \ELSE
\STATE \COMMENT{unable to integrate \pve mode security tasks}	                   \STATE \textbf{return} $\mathsf{Unschedulable}$  
                    \ENDIF
        
    \end{ALC@g}
              
    \STATE \EndFunction
     \\\hrulefill 
  \vspace*{0.5em}

  \STATE 
  \Function{ActiveModeParamSelection($\Gamma_R$, $\Gamma_S^{ac}$, $l_S$)}
  
  \begin{ALC@g}
  \STATE $\mathsf{Schedulable} :=$ \textbf{false}
                    \STATE Initialize \ave mode security task's period $\mathbf{T}(l')_{\forall l' \in [l_S, m]} := [T_i^{des}]_{\forall \tau_i \in \Gamma_S^{ac}}^{\mathsf{T}}$ 
					\FORALL{ priority level $l' \in [l_S, m]$
                    }

					\STATE Solve $\mathbf{P}\mathbf{\ref{opt:server_param_ave}}$ to obtain server parameters
					\IF {$\mathsf{Solution Found}$}
					
                    \STATE Solve $\mathbf{P}\mathbf{\ref{opt:period_adapt_server_ave}}$ to obtain security periods
                    
                    \IF {$\mathsf{Solution Found}$}
					
                    \STATE \COMMENT{store the parameters for priority level $l'$ where $Q^*$, $P^*$ and $\mathbf{T}^*$  are the solutions obtained by $\mathbf{P}\mathbf{\ref{opt:server_param_ave}}$ and $\mathbf{P}\mathbf{\ref{opt:period_adapt_server_ave}}$}
                    \STATE $Q(l'):= Q^*, P(l'):=P^*, \mathbf{T}(l') := \mathbf{T}^*$ 	
                    \STATE $\mathsf{Schedulable} :=$ \textbf{true}
                    \ENDIF

					\ENDIF
					\ENDFOR	
                    \STATE \COMMENT{obtain the parameters that provide best metric}
                    \IF {$\mathsf{Schedulable}$}
                    \STATE Find the priority-level $l^*$ from the solution vector $\mathbf{T}(l')_{\forall l' \in [l_S,m] | \text{~tasks at~} l' \text{~is~} \mathsf{schedulable}}$ that gives the maximum cumulative tightness $\eta^{ac} = \sum_{\tau_i \in \Gamma_S^{ac}} \eta_i$
                    \STATE Set $\mathbf{T}^{\boldsymbol{ac}} := \mathbf{T}(l^*)$, $Q^{ac} := Q(l^*)$, $P^{ac} := P(l^*)$
                    \STATE \COMMENT{return the parameters}
                    \STATE \textbf{return} $l^*$, $\mathbf{T}^{\boldsymbol{ac}}$, $Q^{ac}$, $P^{ac}$
                    
                    \ELSE
\STATE \COMMENT{unable to integrate \ave mode security tasks}	                   \STATE \textbf{return} $\mathsf{Unschedulable}$          
                    \ENDIF
  
 \end{ALC@g}
 
    \STATE 
    \EndFunction
			\end{algorithmic}

			\caption{Feasibility Checking and Parameter Selection}
			\label{alg:sec_schd}
		\end{algorithm}


To find the \pve mode parameters, we initialize the security task's period with the desired period and solve the server parameter selection problem $\mathbf{P}\mathbf{\ref{opt:server_param_pve}}$ (Lines 10--11). If there exists a solution (\eg the constraints are satisfied), we then obtain the periods of the security tasks by solving $\mathbf{P}\mathbf{\ref{opt:period_adapt_server_pve}}$ (Line 13). In the event that neither of these optimization problems returns a solution, we report the task-set as unschedulable (Line 20), since it is not possible to execute security tasks opportunistically without violating real-time constraints.

To select \ave mode parameters, the algorithm iterates through each of the acceptable priority-levels $[l_S,m]$ 
and tries to obtain the periods that maximize tightness for periodic monitoring without violating the real-time constraints (Lines 26--36). 
If there exists a solution (\eg constraints in $\mathbf{P}\mathbf{\ref{opt:server_param_ave}}$ and $\mathbf{P}\mathbf{\ref{opt:period_adapt_server_ave}}$ are mutually consistent), we store the 
solution in a candidate list. The algorithm then finds the best priority-level from the candidate solution sets that provides the maximum tightness (Line 39). 
In the event that no candidate solutions are found for any of the allowable priority ranges, the algorithm reports the task-set as unschedulable.

If \textit{both} the \pve and \ave mode tasks are schedulable, then Algorithm \ref{alg:sec_schd} returns the corresponding periods and the \ave mode priority-level (Line 4). Otherwise, the system is considered as unschedulable (Line 7) since it is not possible to integrate security tasks with desired requirements. This unschedulability result hints that the designers of the system should update system parameters (\eg the number of security tasks, desired and maximum allowable periods of the security tasks, periods of the real-time tasks, if permissible, \etc) in order to integrate security mechanisms.

\section{Evaluation} \label{sec:evaluation}

We evaluate \coolname with randomly generated synthetic workloads 
 (Section \ref{subsec:results}) as well as a proof-of-concept implementation on an ARM-based  embedded development board and real-time Linux (Section \ref{subsec:implementation_BBB}).

\subsection{Experiment with Synthetic Task-sets}
\label{subsec:results}




\subsubsection{Simulation Setup}
\label{subsec:ex_setup}

In order to generate task-sets with an even distribution of tasks, we grouped the real-time and security task-sets by base-utilization from $[0.01+0.1 \cdot i, 0.1+0.1 \cdot i]$, where $i \in \mathbb{Z} \wedge 0 \leq i \leq 9$. Each utilization group contained $500$ task-sets. In other words, a total of $5000$ task-sets were tested for each of the experiments. The utilization of the real-time and security tasks were generated by the UUniFast \cite{uunifast} algorithm and we used GGPLAB \cite{ggplab} to solve the optimization problems.

We used the parameters similar to those used in earlier research \cite{sg1,mhasan_rtss16}. In particular, 
each task-set instance contained $[3, 10]$ real-time and $[2, 5]$  security tasks in each of the modes. Each real-time task $\tau_j \in \Gamma_R$ had a period $T_j \in [10~\rm{ms}, 100~\rm{ms}]$
and we assumed $l_S = \lceil 0.4 m \rceil$.  The desired periods for the security tasks $\forall \tau_i \in \lbrace \Gamma_S^{pa} \cup \Gamma_S^{ac} \rbrace$ were selected from $[1000~\rm{ms}, 3000~\rm{ms}]$ and the maximum allowable period was assumed to be $T_i^{max} = 10 T_i^{des}$.  We considered $\omega_i = 1, ~\forall \tau_i \in \lbrace \Gamma_S^{pa} \cup \Gamma_S^{ac} \rbrace$ and the total utilization of the security tasks was assumed to be no more than $30\%$ of the real-time tasks.  


\subsubsection{Results}

\paragraph{Impact on Cumulative Tightness}

\begin{figure}[!t]
\vspace{-1.00\baselineskip}
\centering
\includegraphics[width=2.9in]{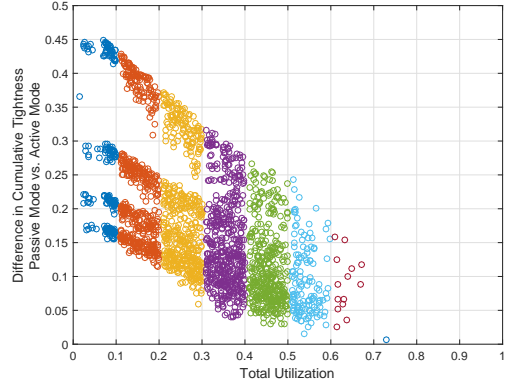}
\caption{\pve mode vs. \ave mode: difference in cumulative tightness of achievable periodic monitoring, $\eta^{av} - \eta^{pa}$. Non-zero difference indicates that the \ave mode tasks achieve better tightness than \pve mode tasks. 
Task-sets from different base-utilization groups are represented by different colors. Each of the data points represents schedulable task-sets.}
\label{fig:diff_eta_active_passive}
\vspace{-0.5\baselineskip}
 \end{figure}

In Fig.~\ref{fig:diff_eta_active_passive} one can see the difference in the tightnesses of the periodic monitoring obtained by \pve and \ave mode (\ie $\eta^{ac} -\eta^{pa}$). For fair comparison we used the same task-sets for both modes. The x-axis of Fig.~\ref{fig:diff_eta_active_passive} represents the total system utilization (\eg utilization of both real-time and security tasks). The positive values in the y-axis of Fig.~\ref{fig:diff_eta_active_passive} imply that the \ave mode tasks obtain better tightness that the \pve mode tasks.  

The figure shows that \ave mode tasks can achieve better cumulative tightness, and that the cumulative tightness $\eta^{pa}$ is comparatively better in low to medium utilization. The main reason is that in \ave mode security tasks are allowed to execute with higher priority, that causes less interference and eventually increases the feasible region in the optimization problems (and hence provides better tightness). For higher utilizations the difference is close to zero. 
This is because, as utilization increases there is less slack in the system, making it difficult to schedule security tasks frequently and resulting in similar levels of tightness for both modes.

\paragraph{Effectiveness of Security}

\begin{figure}[!t]
\centering
\includegraphics[width=2.9in]{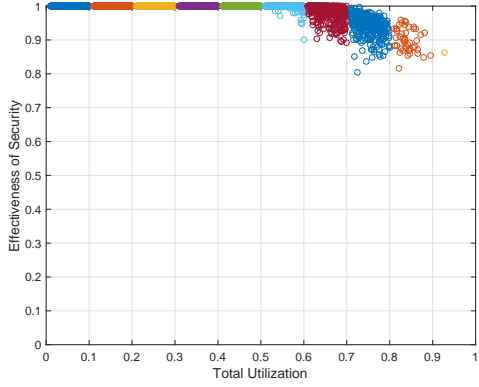}
\caption{The effectiveness of security  vs. total utilization of the system. The closer the y-axis values to $1$, the nearer each security task's period is to the desired period. Task-sets from different base-utilization groups are distinguished by different colors. 
}
\label{fig:effect_security}
 \end{figure}

The parameter $\eta^{(\cdot)}$ is given by the total number of security tasks and provides insights on cumulative measures of security. However, in this experiment (refer to Fig.~\ref{fig:effect_security}) we wanted to measure the effectiveness of the security of the system by observing whether \textit{each} of the security tasks in any mode can achieve an execution frequency closer to the desired one. Hence we used the following metric: 
$\xi = 1- \frac{ {\lVert \mathbf{T}^* - \mathbf{T^{des}}\rVert}_2}{{\lVert \mathbf{T^{max}} - \mathbf{T^{des}}\rVert}_2}$ where $\mathbf{T}^*$ is the solution obtained from Algorithm \ref{alg:sec_schd}, $\mathbf{T^{des}} = [T_i^{des}]_{\forall \tau_i }^{\mathsf{T}} $ and $\mathbf{T^{max}} = [T_i^{max}]_{\forall \tau_i }^{\mathsf{T}} $ are the desired and maximum period vector (refer to Section \ref{subsec:ex_setup}), respectively, and ${\lVert \cdot \rVert}_2$ denotes the Euclidean norm. The closer the value of $\xi$ to $1$, the nearer  each of the security task's  period is to the  desired period.

As the total utilization increases, the feasible set of the period adaptation problem that respects all constraints in  the optimization problems becomes more restrictive. As a result, we see the degradation in effectiveness (in terms of $\xi$) for the task-sets with higher utilization. However,
from our experiments we find that \coolname can achieve periods that are \textit{within $18\%$ of the desired periods}.

\paragraph{Impact on the Schedulability}

\begin{figure}[!t]
\centering
\includegraphics[width=2.9in]{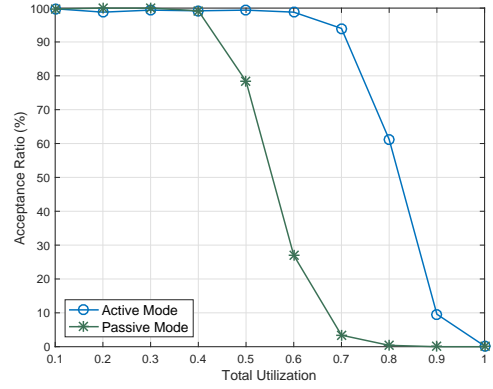}
\caption{Schedulability of real-time and security tasks in both modes. The acceptance ratio is defined by the ratio of the number of accepted
task sets over the total number of generated tasks. For each of the data points, 500 individual task-sets were tested.}
\label{fig:active_passive_sched}
\vspace{-1.0\baselineskip}
\end{figure}

We used the \textit{acceptance ratio}
metric to evaluate schedulability. The acceptance ratio (y-axis in Fig.~\ref{fig:active_passive_sched}) is defined as the number of accepted task-sets (\eg the task-sets that satisfied all the constraints) over the total number of generated ones. As depicted in Fig.~\ref{fig:active_passive_sched} the \ave mode task-set achieves better schedulability compared to the \pve ones. Recall that \ave mode task-sets can be promoted up to priority level $l_S$. As a result \ave mode security tasks potentially experience less interference than the \pve ones. This flexibility gives the optimization routines a larger  feasibility region to satisfy all the constraints. 

\subsection{Experiment with Security Applications in an Embedded Platform} \label{subsec:implementation_BBB}

To observe the performance of the proposed scheme in a practical setup, we implemented \coolname on an embedded platform. 
Our experimental platform \cite{bbb} was configured with 1 GHz ARM Cortex-A8 single-core processor and 512 MB RAM. We used Linux as the operating system -- that allowed us to utilize the existing Linux-based IDSes (refer to Section \ref{subsubsec:sec_task}) for the evaluation. Since the vanilla Linux kernel is  unsuitable for hard real-time scheduling, we enabled the real-time capabilities with the Xenomai \cite{xenomai} 2.6.3 real-time patch (kernel version 3.8.13-r72) on top of an embedded Debian Linux console image. 

We measured the WCET of the real-time and security tasks using ARM cycle counter registers (\eg CCNT), giving us nanosecond-level precision. Since these registers are not enabled by default, we developed a Linux kernel module to access the registers from our application codes.
Our prototype implementation was developed in C and uses a fixed-priority scheduler powered by the Xenomai real-time patch. Sporadic real-time and security tasks in the system were defined by Xenomai $\mathtt{rt\_task\_create()}$ function and were 
suspended after the completion of corresponding instances using the $\mathtt{rt\_task\_wait\_period()}$ function.

\subsubsection{Real-time Tasks}

For a real-time application, we considered a UAV control system (refer to Table \ref{table:rt_task_param}). 
We implemented it using an open-source UAV model \cite{khan-drone}. 
The original application codes were based on the STM32F4  micro-controller (ARM Cortex M4) and developed for FreeRTOS \cite{free_rtos}. Because of differences in library support and execution semantics, we updated the source codes accordingly and ported them to Linux.

\begin{table*}
\caption{Real-time task parameters for the UAV control system 
}
\label{table:rt_task_param}

\centering
\begin{tabular}{P{2.3cm} P{9.50cm} P{1.00cm}}
\hline
Task & Function & Period (ms)\\
\hline
\hline 
Guidance & Select the reference trajectory (\ie altitude and heading)  & 1000 \\ \hline
Controller & Execute closed-loop control functions (\eg actuator commands) & 5000 \\ \hline
Reconnaissance & Read radar/camera data, collect sensitive information and send data to the base control station & 10000 \\
\hline
\end{tabular}

\end{table*}

\subsubsection{Security Tasks}
\label{subsubsec:sec_task}

To integrate security in the aforesaid control system, we included additional security tasks. For the security tasks, we considered two lightweight open-source intrusion detection mechanisms, \ci Tripwire \cite{tripwire}, that detects integrity violations by storing clean system state during initialization and using it later to detect intrusions by comparing the current system state against the stored clean values,  and \cii Bro \cite{bro} that monitors anomalies in network traffic. As Table \ref{table:rtos} shows, 
we consider several security tasks in both modes, \eg  \textit{protecting security task's own binary files}, \textit{protecting system binary and library files}, \textit{monitoring network traffic}.~
In each mode, we set the desired and maximum allowable periods of the security tasks such that utilization of the security tasks did not exceed $50\%$ of the total system utilization. 



\begin{table*}[!htb]
\caption{Security tasks used in the experiments}
\label{table:rtos}

\centering
\begin{tabular}{P{4.50cm} P{6.00cm} P{2.00cm}}
\hline
Task & Function & Mode\\
\hline
\hline 
Check own binary of the security routine (Tripwire) & Scan files (\viz compare their hash value) in the following locations: $\mathtt{/usr/sbin/siggen}$, $\mathtt{/usr/sbin/tripwire}$, $\mathtt{/usr/sbin/twadmin}$, $\mathtt{/usr/sbin/twprint}$,
$\mathtt{/usr/local/bro/bin}$
&  \ave \\ \hline
Check critical executables (Tripwire)& Scan file-system binary ($\mathtt{/bin}$, $\mathtt{/sbin}$) & \ave and \pve  \\ \hline
Check critical libraries (Tripwire) & Scan file-system library ($\mathtt{/lib}$) & \ave   \\ \hline
Monitor network traffic (Bro) & Scan predefined network interface ($\mathtt{en0}$) & \ave and \pve   \\
\hline 
\end{tabular}

\end{table*}


\subsubsection{Experience and Evaluation}

\paragraph{Performance Impact in Different Modes}

 \begin{figure}[!t]
\centering
\includegraphics[width=2.9in]{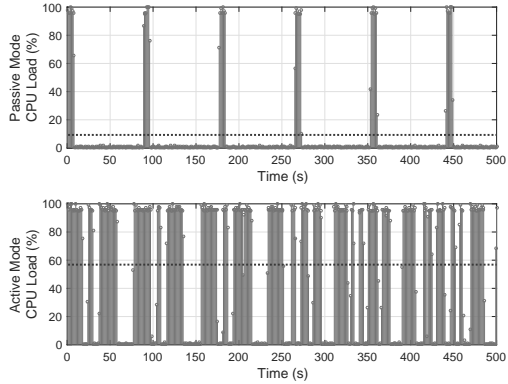}
\caption{The CPU load when the security tasks executed in \pve (top) and \ave (bottom) mode, respectively. The dotted line represents average load over the observation duration (500 s).}
\label{fig:cpu_usage}
\vspace{-1.0\baselineskip}
 \end{figure}

In the first set of experiments, we measured the average CPU load when the security tasks were executing in \pve and \ave modes. For that, we executed the security tasks independently for $500~\mathrm{s}$ in \pve and \ave modes and observed the CPU load using $\mathtt{/proc/stat}$ interface (represents the y-axis of Fig.~\ref{fig:cpu_usage}). As Fig.~\ref{fig:cpu_usage} shows, running security tasks in \ave mode increased the average CPU load compared to running them in \pve mode. This is because \ave mode contains more security tasks (\eg 4 compared to 2, refer to 
Table \ref{table:rtos}) and they execute more frequently than in \pve mode. Because of the nature of applications, most RTS prefer predictability over performance. The overhead of running security tasks in \ave mode comes with increased security guarantees that will suffice for many RTS.


\paragraph{Impact on Detection Time}

To study the detection performance we injected malicious code into the system that mimics anomalous behaviors.  We assumed that an attacker can take over\footnote{One way to override a task could be to use an approach similar to one presented in the literature \cite{cy_side_channel} that exploits the deterministic behavior of the real-time scheduling.} one of the low-priority real-time tasks (referred to as the victim task) and is able to insert malicious code that can execute with a privilege similar to that of legitimate tasks.
We launched the attack at both the \textit{network} and \textit{host}-level. We defined network-level DoS attacks as too many rejected usernames and passwords submitted from a single address and used a real FTP DoS trace \cite{bro_dos} to demonstrate the attack. Malware (such as LRK, tOrn, Adore, \etc) in general-purpose Linux environments causes damage to the system by modifying or overwriting the system binary \cite[Ch. 5]{linux_hack}. Thus we follow a similar approach to demonstrate a host-level attack, \viz we injected ARM shellcode \cite{arm_shellcode} to override the victim task's code and launched the attack by modifying the contents in the file-system binary. 

We obtained the periods of the security tasks in both modes by solving the period adaptation problem (Algorithm \ref{alg:sec_schd}) and set it as the period of security tasks (by using the Xenomai $\mathtt{rt\_task\_set\_periodic()}$ function). For each of the experiments, the work-flow was as follows. We started with a clean (\eg uncompromised) system state, launched the DoS attack at any random time of the program execution and then injected the shellcode after a random interval, and finally logged the time required by security tasks  to detect the attacks. Initially the security tasks ran in \pve mode. When the network-level attack was suspected by the security task (\eg Bro), a mode change request was placed and the control was switched to \ave mode with the corresponding \ave mode tasks (see Table \ref{table:rtos}).  
As mentioned in Section \ref{subsec:timeshield_overview}, our focus is \textit{not} on the effectiveness of a particular IDS here but on the effectiveness of integration of the IDSes into RTS. Therefore  we controlled the experimental environment so that the results were not affected by the false positive/negative rates of the IDS used in the evaluation. In particular, both of the launched attacks were detectable by the respective IDSes used in the evaluation. Detection times were measured using ARM cycle counter registers (CCNT). To ensure the accuracy of the detection time measurements, we disabled all the frequency scaling features in the kernel (by using the $\mathtt{cpufrequtils}$ utility) and allowed the platform to execute with a constant frequency (\eg 1 GHz, the maximum frequency of our experimental platform).

   \begin{figure}[!t]
\centering
\includegraphics[width=2.9in]{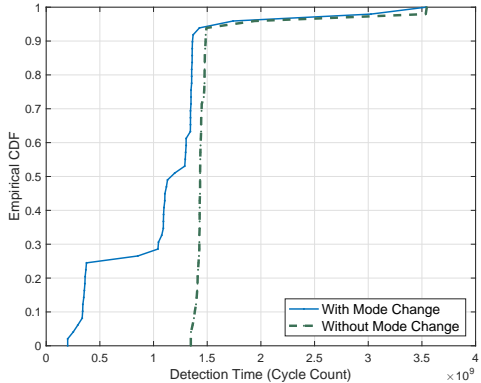}
\caption{The empirical distribution of time to detect the intrusions when mode change was allowed vs when security tasks were run only in \pve mode. We used ARM cycle counter registers to measure the detection time. A total of $50$ individual experiment instances were examined to obtain the timing traces.}
\label{fig:mode_trace}
\vspace{-1.0\baselineskip}
 \end{figure}

We compared the performance of \coolname with that of an earlier approach \cite{mhasan_rtss16} that has no provision for mode changes and in which the security tasks are run with the lowest priority (similar to the \pve mode of operation in \coolname). Specifically, we measured the time to detect both the host and network-level intrusions, and plot the empirical cumulative distribution function (CDF) of those detection times in Fig.~\ref{fig:mode_trace}. The x-axis in  Fig.~\ref{fig:mode_trace} represents the detection time (in cycle count) and the y-axis represents the probability that the attack would be detected by that time.  The empirical CDF is defined as $\widehat{F}_\alpha(\jmath) = \frac{1}{\alpha} \displaystyle \sum_{i=1}^\alpha \mathbb{I}_{[\zeta_i \leq \jmath]}$, where $\alpha$ is the total number of experimental observations, ${\zeta}_i$ is the time taken to detect the attack in the $i$-th experimental observation, and $\jmath$  represents the $x$-axis values (\viz the detection times in cycle count) in Fig.~\ref{fig:mode_trace}. The indicator function $\mathbb{I}_{[\cdot]}$ outputs $1$ if the condition $[\cdot]$ is satisfied and $0$ otherwise.

From Fig.~\ref{fig:mode_trace} we can see that \coolname provides better detection time (\ie fewer cycle counts required to detect the intrusions). From our experiments we find that \textit{on average} \coolname detects attacks $27.29\%$ faster than the reference scheme does. The approach from the literature \cite{mhasan_rtss16} allows the security tasks to run only when other real-time tasks are not running, leading to more interference (\eg higher response times), and does not provide any mechanisms to adapt against abnormal behaviors (\eg the DoS attack in the experiments). In contrast, \coolname allows quick response to anomalies (by switching to \ave mode when a DoS attack is suspected). Since \ave security tasks can run with higher priority and less interference without impacting the timeliness constraints of real-time tasks, \coolname had a superior detection rate in general for most of the experiments  without impacting safety.

\section{Discussion} \label{sec:discussion}

Although \coolname provides an integrated approach to guarantee safety and security in RTS, this framework can be extended in several directions. In the following, we briefly analyze \coolname against different threat models and discuss the limitations of the current framework with possible directions of improvement.

\subsection{Threat Analysis}

The security mechanism will collapse if the adversary can compromise \textit{all} the security tasks. To do so, the adversary would need to intrude into the system, remain undetected and monitor the schedule \cite{cy_side_channel} (to override the security tasks)  \textit{over a long period of time}. Guaranteeing the integrity of the security tasks is an interesting research problem by itself and will be investigated in our future work. While compromising all the security tasks could be \textit{difficult} in practice, it nevertheless would be worthwhile to harden the security posture of \coolname further by \textit{randomizing task schedules} while guaranteeing the safety of the real-time tasks by using approaches similar to one recently proposed in the literature \cite{taskshuffler}. Randomizing the schedule of real-time and security tasks reduces the determinism (and thus the predictability of security tasks' execution) and further reduce the chance of information leakage.  Randomizing task schedules in RTS, unlike traditional systems, is not straightforward since it leads to priority
inversions~\cite{Sha:1990:PIP} that, in turn, cause missed deadlines, and
hence, put the safety of the system at risk.  We intend to incorporate randomization protocols on top of \coolname in future work.

The underlying detection algorithms in security tasks could raise false positive errors that may cause the system to switch modes unnecessarily. Again, a clever adversary may remain undetected and provide a fake indication of malicious activity. This may cause \coolname to frequently switch modes thus reducing performance and availability. Although \coolname \textit{guarantees that the system will remain schedulable} (and hence safe) even with mode changes (refer to Section \ref{subsec:mode_switch_diss}), running of security tasks in \ave mode could impose additional overheads (\ie increased load as we have seen in Fig.~\ref{fig:cpu_usage}) that designers of the system may want to avoid. 

The false positive/negative errors can be mitigated by carefully designing the detection algorithms based on application requirements. Further, we argue that forced mode changes would require an adversary to intrude in the system and \textit{remain undetected for a long time}. In practice that could be \textit{difficult} and \textit{unlikely} in the  presence of several intrusion detection tasks.

\subsection{Limitations and Improvement}

In \coolname each
security task has a desired frequency of execution for better security
coverage. Security tasks so far have been treated as \textit{independent} and
\textit{preemptive}, but in practice, some security monitoring
may need \textit{atomicity} or non-preemptive execution. The server-based model proposed in this work can be extended to incorporate this feature. For example, when a security task needs to perform a special atomic operation, the priority of the server can be increased to a priority that is strictly higher than all (or some) of the real-time tasks. Further, if the security task running under server is not the highest-priority security task, the priority of that task itself will also be increased. If the server's capacity is exhausted while it is executing an atomic operation, we can allow the server to \textit{overrun} \cite{server_overrun}, \ie the server continues to execute at the same priority until the security checking is completed. When the server overruns, the allocated capacity at the start of the next server replenishment period is reduced by the amount of the overrun. However, schedulability analysis must be performed considering maximum blocking times of the security tasks.

Further, security tasks may
have \textit{dependencies} wherein one task depends on the output from one or more other
tasks. For example, an anomaly detection task might depend on the outputs of
multiple scanning tasks, or, the scheduling framework might need to follow
certain \textit{precedence constraints} for security tasks. In
order to ensure the integrity of monitoring security, the security application's
own binary might need to be examined first before it checks the system binary
files. 
In that case, the {\emph cumulative tightness} of the achievable
periodic monitoring proposed in Section~\ref{sec:period_adapt} might no
longer be a reasonable metric. Constraints to ensure that the dependent security tasks are
executed often enough should be included and the optimization problem may need to be
reformulated and evaluated with different metrics.

While \textit{time-to-detect} is a useful metric, it is hard to quantify in a
comprehensive way as it depends on a number of factors such as the efficacy
of monitoring tasks, the kind of intrusion \etc~and is a lagging metric.
Identifying and designing better security metrics is an important and
challenging problem. In future work we will undertake it in the narrow context of
integrating monitoring and detection tasks into RTS.

\section{Related Work}

In our earlier work~\cite{mhasan_rtss16} we proposed to use a server to 
integrate security tasks and execute them opportunistically at a lower priority than real-time tasks. That
approach was useful for legacy RTS
where perturbing 
the schedule of real-time tasks was not an option -- however, the downside was longer time for detection. In contrast, \coolname can respond to anomalous activities in an adaptive manner and provide improved monitoring frequency and detection time when needed. 

A new scheduler \cite{xie2007improving} and enhancements to an existing dynamic priority scheduler \cite{lin2009static} were proposed to meet real-time requirements while maximizing the level of security achieved.
A state cleanup mechanism has been introduced \cite{sg1}, and further generalized \cite{sg2, sibin_RT_security_journal} such that the fixed-priority
scheduling algorithm was modified to mitigate information leakage through shared resources. Researchers  have proposed a schedule obfuscation method \cite{taskshuffler} aimed at randomizing the task schedule while providing the necessary real-time guarantees. Such randomization techniques can improve the security posture by minimizing the predictability of the deterministic RTS scheduler. 
Recent work \cite{slack_cornell, securecore} on dual-core based hardware/software architectural frameworks has aimed to protect RTS against security threats. However, those approaches came at the cost of reduced schedulability or may require architectural/scheduler-level modifications. In comparison, \coolname aims to integrate security \textit{without} any significant modification of the system properties and does \textit{not} violate the temporal constraints or schedulability of the real-time tasks.



Although not in the context of security in RTS, 
there exists other work \cite{delay_period} in which
the authors statically assign the  periods for multiple independent control tasks by considering control delay as a cost metric and estimating the delay through an approximate response time analysis. In contrast, our goal is to ensure security without violating the timing constraints of the real-time tasks. Hence, instead of minimizing response time, we attempt to assign the best possible periods and priority-levels so that we can minimize the perturbation between the  achievable period and desired period for all the security tasks. 

An on-demand fault detection and recovery mechanism has been proposed~\cite{ortega_conf} in which the system can operate in different modes. Specifically, when a fault is detected, a high-assurance
controller is activated to replace the faulty high-performance
controller. While fault-tolerance may also be a design consideration, \coolname  focuses primarily on integrating mechanisms that can foil cyber-attacks. There also exist work in the context of mixed-criticality systems (MCS) where  application tasks of different criticality requirements (\eg deadline and execution time)
share same computation and/or communication resources (refer to literature \cite{mc_review} for a survey of MCS). MCS is different than the problem considered in this work due to the fact that security properties (\ie adaptive switching depending on runtime behavior or frequent execution of monitoring events for faster detection) are often different than temporal requirements (\eg satisfying deadline constraints for mixed-criticality tasks). However, the theory and concepts emerged from MCS can also be applied to the real-time security problems to further harden the security posture of future RTS.






\section{Conclusion}

The sophistication of recent attacks  on UAVs \cite{dronhack},
automobiles \cite{ris_rts_1, checkoway2011comprehensive}, medical devices
\cite{security_medical} as well as an industrial control systems \cite{stuxnet}, indicates that RTS are becoming more vulnerable. In this paper we are making steps towards the development of a comprehensive framework to integrate security mechanisms and  provide a glimpse of \textit{security design metrics} for RTS. Designers of RTS are now able to improve their security posture, which will also improve overall \textit{safety} – and that is essentially the main goal of such systems.

\bibliographystyle{IEEEtran}
\bibliography{references_short}  



\appendix



\section{Solution to the Optimization Problems} \label{appsec:gp}



The \ave and \pve modes parameter selection problems given in Section \ref{sec:sec_server} are constrained nonlinear optimization problems and 
not very straightforward to solve. Therefore we reformulate the optimization problems as a geometric program (GP) \cite{GP_tutorial}. A nonlinear optimization problem can be solved by GP if the problem is formulated as follows: \cite{GP_tutorial}
\begin{subequations}
\begin{align*}
\underset{\mathbf{X}}{\operatorname{min}}~ & f_0(\mathbf{x}),  \\
\text{~Subject to:~} \hspace*{2em} \nonumber \\
 f_i(\mathbf{x}) &\leq  1 \quad i = 1, \cdots, z_p,  \text{~~and~~} \\
 g_i(\mathbf{x}) &=  1 \quad i = 1, \cdots, z_m
\end{align*}
\end{subequations}
where $\mathbf{x} = [x_1, x_2, \cdots, x_z]^{\mathsf{T}}$ denotes the vector of $z$ optimization variables. The functions $f_0(\mathbf{x}), f_1(\mathbf{x}), \cdots, f_{z_p}(\mathbf{x})$ are \textit{posynomial} and $g_1(\mathbf{x}), \cdots, g_{z_m}(\mathbf{x})$ are \textit{monomial} functions, respectively. A monomial function is  expressed as
$
g_i(\mathbf{x}) = c_i \prod\limits_{l = 1}^{L_i} x_l^{a_l},
$
where $c_i \in \mathbb{R}^+$ and $a_l \in \mathbb{R}$. 
A posynomial function (\ie the sum of the monomials) can be represented as
$
f_i(\mathbf{x}) = \sum\limits_{l=1}^{L_i} c_l x_1^{a_{1l}} x_2^{a_{2l}} \cdots x_z^{a_{1l}},
$
where $c_l \in \mathbb{R}^+$ and $a_{jl} \in \mathbb{R}$. 
We can maximize a non-zero posynomial objective function by minimizing its inverse. In addition, we can express the constraint $f(\cdot) < g(\cdot)$ as $\frac{f(\cdot)}{g(\cdot)} \leq 1$.

Based on the above description, we can rewrite the period adaptation problem in either mode as:

\begin{myoptimizationproblem} \label{opt:period_adapt_gp}
\vspace*{-2.0em}
\begin{subequations}
\begin{align*}
\underset{\mathbf{T}^{\boldsymbol{(\cdot)} }}{\operatorname{min}} \sum_{\tau_i \in \Gamma_S^{(\cdot)}} {\omega_i}^{-1}  {(T_i^{des})}^{-1}  T_i \hspace*{-1em} \\
\text{Subject to:~~~} \hspace*{10em} 
\nonumber \\
\Big( \sum_{\tau_i \in \Gamma_S^{(\cdot)}} C_i {T_i}^{-1} \Big) \cdot   \left(n \left[ \left( \tfrac{3 - \tfrac{Q}{P}}{3 - 2 \tfrac{Q}{P}} \right)^{\frac{1}{n}} \hspace*{-1em}- 1\right]\right)^{-1} \hspace*{-1em} &\leq 1\\
(3P^{(\cdot)} - 2Q^{(\cdot)}) {T_i} ^{-1} &\leq 1,  ~\forall \tau_i \in \Gamma_S^{(\cdot)} \hspace*{2em} \\
T_i^{des} {T_i}^{-1} &\leq 1,  ~\forall \tau_i \in \Gamma_S^{(\cdot)}  \hspace*{2em} \\
 {(T_i^{max})}^{-1} T_i &\leq 1,  ~ \forall \tau_i \in \Gamma_S^{(\cdot)} 
\end{align*}
\end{subequations}
\end{myoptimizationproblem}
\hspace{-1.4em} 
where for any symbol $y^{(\cdot)}$ represents the corresponding variable in the representative mode (\eg \pve or \ave).

The above GP formulation $\mathbf{P}\mathbf{\ref{opt:period_adapt_gp}}$ is not a convex optimization problem since the posynomials are not convex functions \cite{GP_tutorial}. However, by using logarithmic transformations (\eg representing $\tilde{T}_i = \log T_i$ and hence $T_i = e^{\tilde{T}_i}$, and replacing inequality constraints of the form $f_i(\cdot) \leq 1$ with $\log f_i(\cdot) \leq 0$), we can convert the above formulation into a convex optimization problem. This convex optimization reformulation can be solved using standard algorithms, such as \textit{interior-point} method in polynomial time \cite[Ch. 11]{boyd_book}.

The server parameter selection problem can also cast into GP as follows. The objective function (\eg the ratio between server capacity and period) can be represented as $Q^{(\cdot)} \left(P^{(\cdot)}\right)^{-1}$, which is clearly a posynomial. The server schedulability constraints (\eg Eq.~(\ref{eq:ser_con1_pve}) and Eq.~(\ref{eq:ser_con1_ave})) can be rewritten as \begin{equation} \label{eq:posy_server1}
\left(Q^{(\cdot)} + \Delta_{S^{(\cdot)}} \right) \left(P^{(\cdot)} \right)^{-1} \leq 1 
\end{equation}
where $\Delta_{S^{(\cdot)}} =
  \sum\limits_{\tau_h \in hp_R(\tau_S^{(\cdot)})} (P^{(\cdot)} + T_h) \cdot T_h^{-1} \cdot C_h $. 
  
  Likewise, the real-time task schedulability constraints for the \ave mode (\eg Eq.~(\ref{eq:ser_con3_ave})) can be represented as
  \begin{equation}
  \begin{aligned}
\Big(  C_j + \hspace*{-1em} \sum_{\tau_h \in hp_R(\tau_j)} \left\lceil \frac{D_j}{T_h} \right\rceil C_h \Big)D_j^{-1} ~~+ \\ 
\left( D_j \left(P^{ac}\right)^{-1} Q^{ac} + Q^{ac} \right) D_j^{-1} \leq 1, ~~\forall \tau_j \in lp_R(\tau_\mathcal{S}^{ac}).
  \end{aligned}
  \end{equation}
 In addition, by using the geometric mean approximation \cite[Ch. 2]{gp_comm} of posynomials we can rewrite the schedulability constraints for the security tasks (\eg Eqs.~(\ref{eq:ser_con2_pve}) and (\ref{eq:ser_con2_ave})) as follows \cite{mhasan_rtss16}:
 \begin{equation}
 \begin{aligned}
\left[P^{(\cdot)}  (Q^{(\cdot)} + I_i^{(\cdot)}) + \Delta_{S^{(\cdot)}}  Q^{(\cdot)} \right] \cdot {\left[Q^{(\cdot)} \cdot  \hat{g}(Q^{(\cdot)}, T_i) \right]}^{-1} \leq 1, \\ 
\forall \tau_i \in \Gamma_S^{(\cdot)} \hspace*{1em}
 \end{aligned}
\end{equation} 
where $\hat{g}(Q^{(\cdot)}, T_i) = {\left( \frac{Q^{(\cdot)}}{\frac{y_0}{y_0 + T_i}} \right)}^{\frac{y_0}{y_0 + T_i}} \cdot {\left( \frac{T_i}{\frac{T_i}{y_0 + T_i}} \right)}^{\frac{T_i}{y_0 + T_i}}$ and $y_0 \in \mathbb{R}^+$ is a given initial point.
  After the logarithmic transformation (\eg $\tilde{Q}_i^{(\cdot)} = \log Q_i^{(\cdot)}$, $\tilde{P}_i^{(\cdot)} = \log P_i^{(\cdot)}$ and replacing the inequality constraints $f_i(\cdot) \leq 1$ with $\log f_i(\cdot) \leq 0$), the objective function and the constraints become a standard convex optimization problem that is solvable in polynomial time.


\end{document}